\documentclass[aps,twocolumn,pra]{revtex4-1}

\usepackage{newlfont}
\usepackage{color}         
\usepackage{soul}            
\usepackage{amssymb}
\usepackage{amsfonts}
\usepackage{amsmath}
\usepackage{wasysym}
\usepackage{graphicx}
\usepackage{epsfig}
\usepackage{amsthm}
\usepackage{bm}
\usepackage{bbold}
\usepackage{mathrsfs}
\usepackage{epstopdf}

\usepackage[colorlinks=true, citecolor=blue, urlcolor=blue ]{hyperref}

\begin{document}

\definecolor{red}{rgb}{1,0,0}
\definecolor{orange}{rgb}{1,0.3,0}

\title{
Subtraction  better than addition: Entanglement in multimode squeezed vacuum \\
post-interface with photons}

\author{Tamoghna Das, R. Prabhu, Aditi Sen(De), and Ujjwal Sen}

\affiliation{Harish-Chandra Research Institute, Chhatnag Road, Jhunsi, Allahabad 211 019, India}

\begin{abstract}
We investigate the entanglement patterns of photon-added and -subtracted
four-mode squeezed vacuum states. Entanglements in different scenarios are
analyzed by varying the number of photons added or subtracted in certain
modes, which are referred to as the ``player'' modes, the others being
``spectators''. We find that the photon-subtracted state can give us
higher entanglement than the photon-added state which is in contrast of
the two-mode situation. We also study the logarithmic negativity of the
two-mode reduced density matrix obtained from the four-mode state which
again shows that the state after photon subtraction can possess higher
entanglement  than that of the photon-added state, and we then compare it
to that of the two-mode squeezed vacuum state. Moreover, we examine the
non-Gaussianity of the photon-added and -subtracted states to find that
the rich features provided by entanglement cannot be captured by the
measure of non-classicality.
%
\end{abstract}

\maketitle

\section{Introduction}
Distribution of entanglement in a multipartite quantum system is known to be a useful resource in several quantum communication and quantum computational tasks \cite{HHHH_RMP}. Notable ones include quantum secret sharing \cite{QCryp}, distributed quantum dense coding \cite{DistriDC}, distribution and concentration of quantum state \cite{Q_state_distri}, and cluster state quantum computing \cite{cluster_comp}. Such protocols have  successfully been realized  in physical systems like photons \cite{Pan10photon}, ions \cite{ION}, nuclear magnetic resonance \cite{NMR}, nitrogen vacancy centers \cite{NVcenter}, etc.  

One of the physical systems in which quantum information tasks have been realized in the laboratory is the class of continuous variable (CV) systems. Historically, the notion of the quantum correlated state of two particles in CV systems first arrived in the seminal paper of Einstein, Podolosky, and Rosen in 1935 \cite{EPR_paper}. In recent years, several communications schemes like teleportation \cite{TeleBennett}  and  classical information transfer by quantum channels \cite{DC1992}, have extensively been investigated both theoretically and experimentally, in CV systems, especially in   Gaussian states \cite{Braun_tele, Holevo_Capacity_CV, Multi_ent_mea_CV,CV_reali}. However, it has been discovered that there are several protocols which can not be implemented using  Gaussian states with Gaussian operations. Examples include entanglement distillation \cite{Ent_dist}, measurement-based universal quantum computation \cite{Measurement_qun_com}, teleportation \cite{NonG_tele}, and quantum error correction \cite{Non_G_error_cor}. The increasing importance of non-Gaussian states have led to the discovery of several mechanisms to create such states in the laboratory \cite{lab_state}. An important one is  adding and subtracting photons, when the initial state is the squeezed vacuum state. Starting with the single mode squeezed vacuum state, whose Wigner function \cite{Wigner} is always positive, it was shown that photon addition can generate a negative dip of the Wigner function in the phase space \cite{R_prabhu_non_c} and hence can deviate from being a Gaussian state. In case of the two-mode squeezed vacuum (TMSV) state  as input state, both entanglement and fidelity of teleportation can be increased by adding and subtracting photons to (from) one or two modes \cite{Cerf_Photon_Add_sub}. For such experiments, see  \cite{lab_demo}. Moreover, the entanglement content of the photon-added state obtained from the TMSV state was shown to be always higher than that of the photon-subtracted state \cite{Cerf_Photon_Add_sub}. 

 Investigations of the squeezed vacuum state with respect to photon addition and subtraction are usually restricted to the two-mode case, even though the importance of multimode CV system is unquestionable. In this paper, we consider the four-mode squeezed vacuum (FMSV) state  as input and deGaussify it by adding and subtracting photons in different modes.   
 We evaluate entanglement between different modes in all possible bipartitions and compare the results of the photon-added state with the subtracted ones. We call a mode as ``spectator'' mode in which either no photon or fixed number of photons are added (subtracted). The other modes are referred to as ``players''. 
We here investigate  two scenarios -- $(1)$ one player mode $(2)$ two player modes.
 We analytically show that in the single player case i.e., when  photons have been added (subtracted) to (from) a single mode, in the player : spectator bipartition, entanglements of the photon-added and -subtracted states coincide. In this situation, we prove that entanglements in both photon-added and -subtracted states monotonically increase with the number of photons added or subtracted. 
  Unlike the TMSV case, we observe that there exists  scenarios in which photon-subtracted output states obtained by subtracting photons from one or two modes contain higher entanglement,  compared to the photon-added state.  
  Specifically, we find that the photon-subtracted state contains more entanglement in the spectator : rest bipartition than that of the photon-added state, when a single mode acts as a player. Similar hierarchy can also be obtained when any two modes act as players. Interestingly, the advantageous situation for photon subtraction can be reversed by adding fixed number of photons in the spectator modes.
 Such behavior can also be  viewed by analyzing logarithmic negativity of the output two party state which can be obtained by discarding either two player modes or two spectator modes. Finally, we study a distance-based measure of non-Gaussianity in these scenarios and find that the non-Gaussianity in general is higher for the photon-added state than that of the photon-subtracted state. In case of two modes, photon-added states are known to be more non-Gaussian than the photon-subtracted states. However,  as already noticed in  \cite{Cerf_Photon_Add_sub}, the relation between entanglement and non-Gaussianity is not straightforward. In the four-mode case, we again find that the photon-added state has always higher non-Gaussianity than that of the photon-subtracted states, and hence reflects that entanglement and non-Gaussianity are possibly not directly connected. 
%

The paper is organized in the following way. In Sec. \ref{Sec:N_mode_squeezedvac}, we discuss briefly the N-mode squeezed vacuum state, and two special cases, the two mode squeezed vacuum state, in \ref{Sec:TMSV_def}, and the four-mode squeezed vacuum state, in \ref{Sec:FMSV_def}. In Sec. \ref{Sec:Photon_Add_Sub_FMSV}, we consider the FMSV state, when $m_i$ number of photons are added to or subtracted from the  mode $i$. In  Secs. \ref{Sec:Non_G} and \ref{Sec:Quantum_correlation_measure}, we briefly introduce a non-classicality measure of a quantum state in a continuous variable (CV) system, and  the quantum correlation measures  which are relevant in the paper, respectively.   In Sec. \ref{Sec:vonNeu_Comparison}, we present the main results  in which we systematically compare the entanglement of the four-mode photon-added state with that of the photon-subtracted state by considering the von Neumann entropy in different bipartitions. Another entanglement measure, the logarithmic negativity, for the photon-added and -subtracted states are evaluated  in Sec. \ref{Sec:LN_Comparison}, while the behavior of non-Gaussianity for the output state is studied 
in Sec. \ref{Sec:NonG_Comparison}. We summarize in Sec. \ref{Sec:Conclusion}. 

\section{N-mode squeezed vacuum state}\label{Sec:N_mode_squeezedvac}
In this section, we discuss the N-mode squeezed vacuum state (NMSV), specifically the two-mode and four-mode squeezed vacuum states, and a state obtained after adding (subtracting) an arbitrary number of photons at the mode $i$. These states are examples of entangled states in continuous variables which can be used in various quantum information tasks. To define such states, let us first denote the bosonic creation and annihilation operators at the mode $i$,  as $\hat{a}_i^{\dagger}$ and {$\hat{a}_i$} respectively, which satisfy the bosonic commutation relations, $[\hat{a}_i,\hat{a}^{\dagger}_j] = \delta_{ij}$, and $[\hat{a}_i,\hat{a}_j] = 0$, $[\hat{a}_i^{\dagger},\hat{a}_j^{\dagger}] = 0$. 
By using bosonic operators, an $N$-mode squeezing operator can be defined as \begin{equation}
{\cal S}(\epsilon) = \exp\left(\frac 12 \sum_{i=1}^N\left(\epsilon^* \hat{a}_i \hat{a}_{i+1} - \epsilon  \hat{a}_i^{\dagger} \hat{a}_{i+1}^{\dagger}\right)\right),
\end{equation} 
where $\hat{a}_{N+1} = \hat{a}_1$. The corresponding (NMSV) state is given by
\begin{eqnarray}\label{Eq:Nmodesq_def}
|\psi_N\rangle &=& {\cal S}(\epsilon)|0_1 0_2 \ldots 0_N\rangle \nonumber \\
&=& \frac{1}{N_S}\exp\left(-\frac{1}{2}\sum_{j,k = 1}^N\hat{a}_j^{\dagger}\tanh(rQ)_{jk}\hat{a}_k^{\dagger} e^{i\theta}\right) \nonumber \\ && \hspace{1.3in} |0_1 0_2\ldots 0_N\rangle,
\end{eqnarray}
where $|0_1 0_2 \ldots  0_N \rangle$ is the N-mode vacuum state, $N_S$, is a normalization constant, and $\epsilon = re^{i\theta}$, with $r$ being the squeezing parameter. Here the matrix $Q$ is obtained from the following relation
\begin{eqnarray}\label{Eq:Q_matrix_def}
{\cal S}(\epsilon)^{\dagger}\hat{a}_j{\cal S}(\epsilon) &=& \sum_k^N (\cosh(rQ)_{jk}\hat{a}_k - \sinh(rQ)_{jk}e^{i\theta}\hat{a}_k^{\dagger}) \nonumber \\
&& \hspace{9em} \forall \hspace{1em} j = 1,\ldots, N.
\end{eqnarray}
Let us now define the position and momentum operators for each mode, given by
\begin{equation}\label{Eq:Cannonical_operator_position_def}
q_i = (\hat{a}_i +\hat{a}_i^{\dagger}),
\end{equation}
\begin{equation}\label{Eq:Cannonical_operator_momentum_def}
p_i = \frac{1}{i}(\hat{a}_i -\hat{a}_i^{\dagger}),
\end{equation}
to show that the  Eq. (\ref{Eq:Nmodesq_def}) indeed represents a squeezed state. 
The variances of the N-mode quadrature operators, $X_1 = \frac{1}{2\sqrt{N}}\sum_j(\hat{a}_j + \hat{a}_j^{\dagger})$ and $X_2 = \frac{1}{2i\sqrt{N}}\sum_j(\hat{a}_j - \hat{a}_j^{\dagger})$, are given by 
\begin{equation}
\Delta X_1^2 = \frac{1}{4}(e^{2r}\sin^2(\theta / 2)+e^{-2r}\cos^2(\theta / 2)),
\end{equation}
and
\begin{equation}
\Delta X_2^2 = \frac{1}{4}(e^{2r}\cos^2(\theta / 2)+e^{-2r}\sin^2(\theta / 2)). 
\end{equation}
Thus for $\theta = 0 $ or $ \pi$, we have $\Delta X_1 \Delta X_2 = \frac{1}{4}$. However, for anyone of the $i = 1,2,$   $\Delta X_i \leq \frac{1}{2}$, while $\Delta X_i \geq \frac{1}{2}$  for  the other $i$.  This guarantees that the state, given in Eq. (\ref{Eq:Nmodesq_def}) is a squeezed state. We assume $\theta = 0 $ throughout the paper.

\subsection{Two mode squeezed vacuum state}\label{Sec:TMSV_def}
The two mode squeezed vacuum state can be obtained by putting $N= 2$ in Eq. (\ref{Eq:Nmodesq_def}), where 
$
Q = \left( {\begin{array}{cc}
0 & 1\\
1 & 0\\
\end{array}} \right)$, and 
\begin{equation}
\tanh rQ = \left( {\begin{array}{cc}
0 & \tanh r\\
\tanh r & 0\\
\end{array}} \right).
\end{equation}
 Thus, the TMSV state with $\theta  = 0$ is given by
 \begin{eqnarray}\label{Eq:TMSV_def}
 |\psi_2 \rangle &=& \text{sech} \,\, r e^{-\tanh r\hat{a}_1^{\dagger}\hat{a}_2^{\dagger}}|00\rangle \nonumber \\
 & = &\text{sech}\,\, r\sum_{n=0}^{\infty}(-\tanh r)^n|n\rangle |n\rangle,
 \end{eqnarray}
where $|n\rangle = \frac{(\hat{a}^{\dagger})^n}{\sqrt{n}}|0\rangle$, is the occupation number state. 

Taking $|\psi_2\rangle$ as the initial state, the behavior of entanglement and non-Gaussianity after adding or subtracting photons, have extensively been investigated \cite{Cerf_Photon_Add_sub}.


\subsection{Four-mode squeezed vacuum state}\label{Sec:FMSV_def}
Let us now consider the FMSV state obtained  by setting $N = 4$ in Eq. (\ref{Eq:Nmodesq_def}). The $4\times 4$ matrix, $Q$, in this case, takes the form
\begin{equation}
Q =\frac{1}{2} \left( {\begin{array}{cccc}
0 & 1 & 0 & 1\\
1 & 0 & 1 & 0\\
0 & 1 & 0 & 1\\
1 & 0 & 1 & 0\\
\end{array}} \right).
\vspace{1em}
\end{equation}
The FMSV state with $\theta = 0$, is then given by \cite{Ma_Rhodes}
\begin{equation}\label{Eq:FMSV_def}
|\psi_4 \rangle = \frac{1}{\cosh r}e^{-\frac{\tanh r}{2}(\hat{a}_1^{\dagger}+\hat{a}_3^{\dagger})(\hat{a}_2^{\dagger}+\hat{a}_4^{\dagger})}|0000\rangle.
\end{equation} 
Expanding the exponential in Eq. (\ref{Eq:FMSV_def}), we have
\begin{eqnarray}\label{Eq:FMSV_expansion}
|\psi_4 \rangle &=& \frac{1}{\cosh r}\sum_{n = 0}^{\infty} \left(-\frac{\tanh r}{2}\right)^n \sum_{r_1,r_2 = 0}^{n}\sqrt{{n \choose r_1}{n \choose r_2}}\nonumber \\
&& \hspace{7em} |n-r_1\rangle|n-r_2\rangle|r_1\rangle|r_2\rangle .
\end{eqnarray}

\subsubsection{Photon-added and -subtracted four-mode state}\label{Sec:Photon_Add_Sub_FMSV}
In this paper, we consider the FMSV state, $|\psi_4\rangle$, as an initial state and our aim is to find the characteristics of its entanglement and the measure of non-Gaussianity after adding and subtracting a finite number of photons. Suppose $m_i$ number of photons are added at each  mode $i$, with $i = 1,2,3,4.$  Then the output four-mode (FM) state reads as 

\begin{widetext}
\begin{eqnarray}\label{Eq:Photon_Add_def}
|\psi^{add}_{\{m_i\}}\rangle &=& \frac{1}{N^{add}}\sum_{n = 0}^{\infty} \left(-\frac{\tanh r}{2}\right)^n \sum_{r_1,r_2 = 0}^{n}\sqrt{{n \choose r_1}{n \choose r_2}}
 \sqrt{\frac{(n-r_1+m_1)!}{(n-r_1)!}}\sqrt{\frac{(n-r_2+m_2)!}{(n-r_2)!}} \sqrt{\frac{(r_1+m_3)!}{r_1!}} \sqrt{\frac{(r_2+m_4)!}{r_2!}} \nonumber \\
&& \hspace{9cm} \times |n-r_1 + m_1\rangle |n-r_2 + m_2\rangle
 |r_1+ m_3\rangle |r_2+m_4\rangle \nonumber \\
 &\equiv& \sum_{n=0}^{\infty}\sum_{r_1,r_2=0}^n p^{\{m_i\}}_{n,r_1,r_2} |n-r_1 + m_1\rangle |n-r_2 + m_2\rangle
 |r_1+ m_3\rangle |r_2+m_4\rangle, 
\end{eqnarray}  
where $N^{add}$ is the normalization constant. Similarly, 
after subtracting $\{m_i\} (i=1,2,3,4)$ number of photons from each mode of the FMSV state, the resulting state is given by
\begin{eqnarray}\label{Eq:Photon_Sub_def}
|\psi^{sub}_{\{m_i\}}\rangle &=& \frac{1}{N^{sub}}\sum_{n = M}^{\infty} \left(-\frac{\tanh r}{2}\right)^n \sum_{r_1 = m_3}^{n-m_1}\sum_{r_2 = m_4}^{n-m_2}
\sqrt{{n \choose r_1}{n \choose r_2}}\sqrt{\frac{(n-r_1)!}{(n-r_1-m_1)!}}\sqrt{\frac{(n-r_2)!}{(n-r_2-m_2)!}}  
 \sqrt{\frac{r_1!}{(r_1-m_3)!}}  \sqrt{\frac{r_2!}{(r_2-m_4)!}} \nonumber\\
&& \hspace{9cm} \times   |n-r_1 - m_1\rangle |n-r_2 - m_2\rangle
 |r_1- m_3\rangle |r_2-m_4\rangle  \nonumber \\
&\equiv & \sum_{n=M}^{\infty}\sum_{r_1=m_3}^{n-m_1} \sum_{r_2=m_4}^{n-m_2} q^{\{m_i\}}_{n,r_1,r_2} |n-r_1 - m_1\rangle |n-r_2 - m_2\rangle  |r_1- m_3\rangle 
|r_2-m_4\rangle, 
\end{eqnarray}
\end{widetext}
where $N^{sub}$ is the normalization constant, and $M = \max \{m_1+m_3, m_2+m_4\}$.

\section{Measure of non-classicality in continuous variable systems }
\label{Sec:Non_G}

The negative  Wigner function of a given state indicates the non-classical nature of the corresponding state while the positivity implies the opposite. On the other hand, it is known that the Wigner function of a Gaussian state is always positive \cite{Optics_book}. Therefore, one can define a measure of non-Gaussianity or non-classicality by measuring the departure of a given state, $\rho$, in a CV system from a Gaussian state. In terms of relative entropy distance, it is given by \cite{Non_G_Genoni,Quantify_non_G_Marco_Genoni,Quantify_non_G_Marian} 
\begin{eqnarray}\label{Eq:non_G_def}
\delta_{NG}(\varrho) &=& S(\varrho||\varrho_G)\nonumber \\
&=& S(\varrho_G)-S(\varrho),
\end{eqnarray}
where $S(\eta||\sigma) = -\text{tr}(\eta \log_2 \sigma) - S(\eta) $, and $\rho_G$ is a Gaussian state which has same covariance matrix and first moment as $\rho$. Here, $S(\sigma) = - \text{tr}(\sigma \log_2\sigma)$ is the von Neumann entropy of $\sigma$. 
 
The von Neumann entropy, $S(\varrho_G)$, of any Gaussian state can be calculated by using its covariance matrix, $\sigma$. For an $N$ mode Gaussian state, $\varrho_G$, the von Neumann entropy is defined \cite{Holevo_Capacity_CV} as
\begin{equation}
S(\varrho_G) = \sum_{k = 1}^N g(\nu_k),
\label{Eq:Entropy_gausianstate}
\end{equation} 
where $\nu_k$ is the Williamson normal form of the covariance matrix of the $N$-mode Gaussian state $\varrho_G$, and the function $g(x)$ is given by
\begin{equation}\label{Eq:Entropy_gaussian_Williamson}
g(x) = -\frac{x+1}{2}\log_2\Big(\frac{x+1}{2}\Big)-\frac{x-1}{2}\log_2\Big(\frac{x-1}{2}\Big).
\end{equation} 

In this paper, we will take a Gaussian state as the input state, and after photon addition (subtraction), the diversion of the output state from the input Gaussian state will be quantified by $\delta_{NG}$.

\section{Quantum correlation measures}
\label{Sec:Quantum_correlation_measure}
Quantum correlation measures in bipartite systems, especially for two qubit systems, are well understood. Such quantifications include the von Neumann entropy of local density matrices for pure states \cite{Ent_entropy},  entanglement of formation \cite{distillable}, concurrence \cite{concEof}, logarithmic negativity \cite{LN}. 
However, measures of quantum correlation in a multipartite scenario, both in discrete and CV systems are limited \cite{HHHH_RMP,Multi_ent_mea_CV}. To characterize entanglement in CV system with multiple modes, one possibility is to compute von-Neumann entropy in different bipartition of modes.
Another possibility is to study logarithmic negativity of two modes which can be obtained after discarding all the modes except two.  
In this section, we briefly discuss the local von Neumann entropy and the logarithmic negativity  in CV systems.

\subsection{Entanglement of a pure state}\label{Sec:von_Neumann_entropy}
Entanglement of a bipartite pure state, can be  defined by the von-Neumann entropy of the reduced density  matrix of a given state \cite{Ent_entropy}, 
i.e.,
\begin{equation}
 E(|\psi\rangle_{AB}) = S (\rho_A)
\end{equation}
where $\rho_A = \text{tr}_B(|\psi\rangle_{AB}\langle\psi|)$. 
 In CV systems, entanglement of a two-mode pure state can  be quantified by the von Neumann entropy of a single mode. The single mode density matrix can be a matrix of infinite dimension which has to be diagonalized to evaluate its von Neumann entropy. The calculation of the entropy can be carried out after truncating the matrix to a large block. 
The block size is determined by checking for convergence of trace, with increasing block-size, to unity up to a certain significant digit. We will discuss this issue in detail for a specific scenario. 

In the multipartite domain, entanglement is difficult to characterize even for  pure states \cite{HHHH_RMP}. However, if one divides a multipartite  system into two blocks, then the entanglement between the two subsystems is the von Neumann entropy of one of the block, provided the system is in a pure state. The entanglement between two blocks of a multiparty state can capture entanglement distribution in the multipartite domain. Such a quantification  has been extensively used in many-body systems \cite{Plenio_RMP}. Here, we divide the multimode system into two parts and investigate the behavior of the entanglement content in the bipartition by adding (subtracting) photons in various modes.

%

\subsection{Logarithmic negativity}\label{sec:log_neg}

In CV systems, logarithmic negativity (LN) is an important entanglement measure \cite{LN}. 
For a state, $\rho_{N}$, with $ N = N_1 + N_2$ modes, it is given by
\begin{eqnarray}
LN(\rho_{N}) = \log_2{\cal N}(\rho_{N}),
\end{eqnarray}
where the negativity of the given state is given by 
\begin{equation}
{\cal N}(\rho_{N}) = 1 + 2 \Big|\sum_i \mu_i\Big|.
\end{equation} 
Here $\mu_i$'s are the negative eigenvalues of the partially transposed density matrix, $\rho^{T_{N_1}}_{N}$, where partial transposition is taken with respect to the $N_1$ modes \cite{PTranspose}. As mentioned for the evaluation of the von Neumann entropy, LN is also calculated by truncating to  a large block of the infinite dimensional matrix. 


\section{Comparison of entanglement enhancement between photon addition and subtraction}
\label{Sec:vonNeu_Comparison}

In this section, our aim is to investigate the effects on entanglement in  different bipartitions, when photons are added (subtracted) in (from) different modes of a four-mode squeezed vacuum state. To study such behavior, we divide the modes into two different categories, viz. $(1)$ player modes -- the modes in which number of photons that we add (subtract) varies, and $(2)$ spectator modes -- the  modes in which either no photon or fixed number of photons are added (subtracted) and hence plays a spectator role in the deGaussification process. The comparison has been made between the situations, when the $m_i, i=1,2,\ldots $  photons are added in the player modes, and the scenario  when the same number of photons are subtracted from the player modes. To execute such comparison, we introduce a quantity 
\begin{eqnarray}
\label{Eq:delta_E}
\delta^E_{\cal A}(\{m_i\}) &=&  E(\rho_{\cal A:B }^{add\{m_i\}}) -  E(\rho_{\cal A:B}^{sub\{m_i\}}) 
\end{eqnarray}
where ${\cal A:B}$ is a bipartition with ${\cal A} \cap {\cal B} = \emptyset$. 
 The positivity of $\delta^E(\{m_i\})$ implies that addition is better than  subtraction from an entanglement perspective. It is clear that the behavior of $\delta^E_{\cal A}(\{m_i\})$
with $\{m_i\}$ depends on the number of player and spectator modes as well as the bipartite splits.



\subsection{Photon added and subtracted with one player mode }
\label{SubSec:oneplayer}

\begin{figure}[t]
\begin{center}
 \includegraphics[width=0.95\columnwidth,keepaspectratio,angle=0]{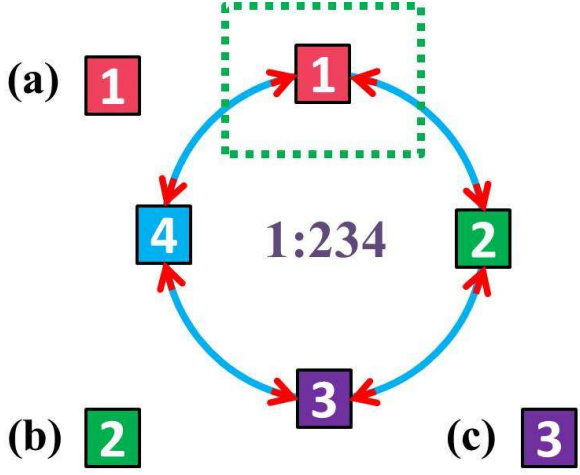}
 \end{center}
\caption{(Color Online) Schematic diagram of choices of player and spectator modes as well
as partitions. If we fix the bipartition  to be $1:234$ there are three nontrivial possibilities of choosing a single player in the photon-added and the subtracted FM state. There are the cases (a) - (c), and the number in the square mentioned for each case is the mode at which the photon is added/subtracted. }
\label{fig:Depict_1}
 \end{figure}

Let us first consider a situation in which one mode acts as a player while the rest are the spectator modes. We first restrict ourselves in the $1:234$ cut irrespective of the choice of the player mode. In this case, there exists three different possibilities of choosing a player mode $(a)$ first mode as player and the rest as spectators, $(b)$ second mode as player and $(c)$ third as player (see Fig. \ref{fig:Depict_1}). From Eqs. (\ref{Eq:Photon_Add_def}) and (\ref{Eq:Photon_Sub_def}), it is clear that fourth mode as a player is equivalent with case $(b)$, and hence we exclude this case.

\subsubsection{Single  player mode in the smallest bipartition}

Suppose that we add or subtract $m_1$  photons in the first mode without putting any number of photons in rest of the modes, as shown in Fig. \ref{fig:Depict_1}(a). Here the first mode acts as a player. The reduced density matrices  can be calculated from Eqs. (\ref{Eq:Photon_Add_def}) and (\ref{Eq:Photon_Sub_def}), which read  as
\begin{eqnarray}\label{Eq:rho_1_m1_add}
\rho_{1,m_1}^{add} &=& \frac{1}{N_1^{add}} \sum_{n=0}^{\infty}\frac{\tanh^{2n} r}{2^n}\sum_{r_1 = 0}^n {n \choose r_1} \frac{(n+m_1-r_1)!}{(n-r_1)!}\nonumber \\
&& \hspace{6em} |n+m_1-r_1\rangle \langle n+m_1-r_1|
\end{eqnarray} 
for photon addition,
and 
\begin{eqnarray}\label{Eq:rho_1_m1_sub}
\rho_{1,m_1}^{sub} 
& = & \frac{1}{N_1^{sub}} \sum_{n=0}^{\infty} \frac{\tanh^{2n} r}{2^n} \sum_{r_1 = 0}^n \frac{(m_1+r_1)!}{r_1!}  {n+m_1 \choose r_1+m_1} \nonumber \\
&& \hspace{14em} |r_1\rangle \langle r_1|
\end{eqnarray} 
for photon subtraction.
 We now analytically establish that entanglement in the bipartition of the player and the spectator modes, increases with the number of photons added. 

{\bf Proposition:} \textit{ Entanglement increases with the addition of a single photon in a four-mode photon-added state, i.e., }
\begin{eqnarray} 
E(|\psi^{add}_{m_1+1}\rangle)_{1:234} &\geq & E(|\psi^{add}_{m_1}\rangle)_{1:234} 
\end{eqnarray}
\textit{where $|\psi^{add}_{m_1 + i}\rangle, i=0, 1$ denotes the state in which $m_1+i$ number of photons are added at the mode $1$.} 

\texttt{Proof:} To evaluate entanglement in the $1:234$ bipartition, we have to study the single mode reduced density matrix, $\rho^{add}_{1,m_1}$, of the four-mode state $|\psi^{add}_{m_1}\rangle$. To prove $E(|\psi^{add}_{m_1+1}\rangle)_{1:234} \geq  E(|\psi^{add}_{m_1}\rangle)_{1:234} 
$, it is equivalent to show $ S(\rho^{add}_{1,m_1+1}) \geq  S(\rho^{add}_{1,m_1})$. After inserting the normalization constant in Eq. (\ref{Eq:rho_1_m1_add}), we get 
\begin{eqnarray}
\rho_{1,m_1}^{add}&=& 2^{m_1}\frac{(1-x)^{m_1 + 1}}{(2-x)^{m_1}} \sum_{r_1=0}^{\infty} f(r_1,x){m_1+r_1 \choose m_1} \nonumber \\
&& \hspace{8em} |m_1+r_1\rangle \langle m_1+r_1| \nonumber \\
&=& \sum_{r_1=0}^{\infty} g(x,m_1,r_1)|m_1+r_1\rangle \langle m_1+r_1|
\end{eqnarray}
where  $x = \tanh^2 r$, 


\begin{equation}
f(r,x) = \sum_{n=r}^{\infty} \frac{x^{n}}{2^n} {n \choose r},
\end{equation}
and
\begin{equation}
g(x,m,r) = 2^{m}\frac{(1-x)^{m + 1}}{(2-x)^{m}} f(r,x){m+r \choose m}. 
\end{equation}
Therefore, entanglement in the player : spectator bipartition is given by
$E(|\psi^{add}_{m_1}\rangle_{1:rest}) =  S(\rho_{1,m_1}^{add}) = -\sum_{r_1 = 0}^{\infty} g(x,m_1,r_1)\log_2g(x,m_1,r_1)$.

Now if we add one more photon to the state in Eq. (\ref{Eq:rho_1_m1_add}), the entanglement is going to be $E(|\psi^{add}_{m_1 + 1}\rangle_{1:rest}) = -\sum_{r_1 = 0}^{\infty} g(x,m_1+1,r_1)\log_2g(x,m_1+1,r_1)$. Let us now evaluate $g(x,m_1+1,r_1)$. 
It simplifies as
\begin{eqnarray}
g(x,m_1+1,r_1)
&=& \frac{2(1-x)}{(2-x)}g(x,m_1,r_1)  \nonumber \\
&&+ \frac{x}{2-x}g(x,m_1+1,r_1-1), 
\end{eqnarray}
by using Pascal's identity, and the recursion relation of $f(r,x)$, which is given by
\begin{equation}\label{Eq:recursion}
f(r,x) = \frac{x}{2-x}f(r-1, x).
\end{equation}
Using the concavity of the function $h(x) = -x\log_2x$, we get
\begin{eqnarray}
h(g(x,m_1+1,r_1))&\geq & \frac{2(1-x)}{(2-x)}h(g(x,m_1,r_1)) + \nonumber \\ && \hspace{-1em} \frac{x}{2-x}h(g(x,m_1+1,r_1-1)).
\end{eqnarray}

Taking the sum over $r_1$ in both sides, we have
\begin{equation}
S(\rho_{1,m_1+1}^{add}) \geq  \frac{2(1-x)}{(2-x)}S(\rho_{1,m_1}^{add}) + \frac{x}{2-x}S(\rho_{1,m_1+1}^{add})
\end{equation}
which immediately implies
\begin{equation}
S(\rho_{1,m_1+1}^{add}) \geq  S(\rho_{1,m_1}^{add}).
\end{equation}
Hence the proof. \hfill $ \blacksquare $
 
Similarly one can also show that entanglement of the photon-subtracted state in the player : spectator split increases with number of photons subtracted from the state.

We are now going to analyze the effects on entanglement under addition and subtraction of same number of photons .

{\bf Proposition:} \textit{When a single mode acts as a player, entanglement between the player and the spectator modes of the photon-added state coincide with that of the photon-subtracted state.} 

\texttt{Proof:} To prove that the increase of entanglement in the multimode state is same for addition and subtraction, we consider the single mode reduced density matrix.  The single site reduced density matrix of photon-subtracted state after inserting $N_1^{sub}$, is given by
\begin{eqnarray}
\label{Eq:rho_1_m1_sub2}
\rho_{1,m_1}^{sub} &=& (1-x)^{m_1+1}\sum_{r_1 = 0}^{\infty} {m_1+r_1 \choose m_1} \nonumber \\ && \hspace{5em} \times \underbrace{\sum_{n = r_1}^{\infty}\frac{x^n}{2^n}{n+m_1 \choose r_1+m_1}}_{f_{sub}(r_1,m_1,x)}   |r_1\rangle\langle r_1|,\,\,\,\,
\end{eqnarray}
where 
\begin{equation}
f_{sub}(r_1,m_1,x) = \frac{x^{r_1} 2^{m_1+1}}{(2-x)^{r_1+m_1+1}},
\end{equation}
which can be obtained by a recursion relation similar to that  given in Eq. (\ref{Eq:recursion}).
On the other hand, the reduced density matrix after adding same number of photons reads as
\begin{eqnarray}\label{Eq:rho_1_m1_add2}
\rho_{1,m_1}^{add} = \sum_{r_1 = 0}^{\infty} 2^{m_1}\frac{(1-x)^{m_1 + 1}}{(2-x)^{m_1}}\frac{2}{2-x}\Big(\frac{x}{2-x}\Big)^{r_1}\nonumber\\
\times {m_1+r_1 \choose m_1} |r_1+m_1\rangle\langle r_1+m_1|.
\end{eqnarray}
Comparing Eqs. (\ref{Eq:rho_1_m1_sub2}) and (\ref{Eq:rho_1_m1_add2}), we have $S(\rho_{1,m_1}^{add}) = S(\rho_{1,m_1}^{sub})$.
 \hfill $ \blacksquare $

\begin{figure}[h]
\begin{center}
 \includegraphics[width=1\columnwidth,keepaspectratio,angle=0]{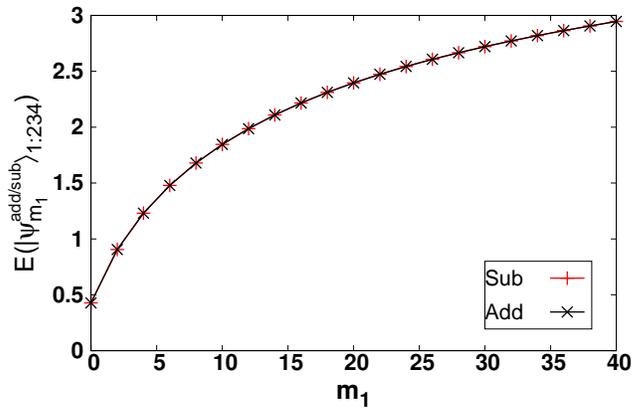}
 \caption{(Color online) Behavior of $E(|\psi_{m_1}^{add}\rangle_{1:234})$ and $E(|\psi_{m_1}^{sub}\rangle_{1:234})$ vs. $m_1$. We add ($\times$) and subtract ($ + $) upto $40$ photons in (from) the first mode, and calculate entanglement  in the $1:234$ bipartition, when no photons are added (subtracted) in (from) the spectator modes. 
 As shown in the propositions, entanglement in both the cases increases monotonically with $m_1$ and they coincide.}
 \label{fig:entropy_1st_1_234} 
 \end{center}
 \end{figure}

To visualize the above Propositions, we plot $S(\rho_{1,m_1}^{add/sub})$, with respect to $m_1$ by fixing the squeezing parameter $r = 0.4$ in  Fig. \ref{fig:entropy_1st_1_234}. It clearly shows that  the curve for photon addition merges with the curve of photon subtraction. Moreover, it shows that entanglement in that bipartition monotonically increases with the addition or subtraction of photons as shown in  Proposition 1. Note here that although the results presented here are when the photons are added at the mode 1 and the bipartition is considered as player : spectator mode, the Propositions remain unaltered if another mode also acts as a player by keeping the similar bipartition.

\subsubsection{Effects on entanglement due to change of partition}

%
We now consider  the entanglement  in the same bipartition as in the previous case, i.e., $1:234$. However,  the second or third mode now act as  player and no photons are added in the rest of the  modes.  
In the previous case, one block contained only the player mode while the other one contains all the spectator modes. In this case, one part of the partition contains one spectator mode while the other one consists of both 
the player and the rest of the spectator  modes. In the previous case, we have already shown that the effects on entanglement due to addition and subtraction of photons are similar. We will now show whether such observation remains
invariant even in this scenario.

\begin{figure}[t]
\begin{center}
 \includegraphics[width=1.05\columnwidth,keepaspectratio,angle=0]{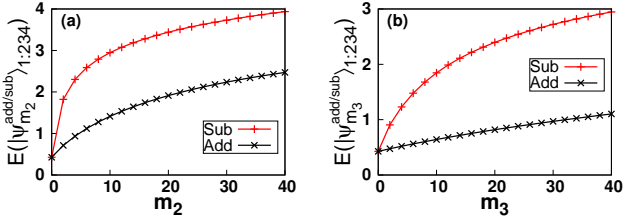}
 \caption{(Color online) (a) Trends of $E(|\psi_{m_2}^{add}\rangle_{1:234})$ and $E(|\psi_{m_2}^{sub}\rangle_{1:234})$ with the number of photon-added (subtracted) in (from) the second mode. (b) Similar study has been carried out when the third mode acts as a player. Both the cases reveal that subtraction is better than addition. }
 \label{fig:entropy_2nd_3rd_1_234}
 \end{center}
 \end{figure}

Let us now take the four-mode squeezed vacuum state as input, and add (subtract) $m_2$ photons in (from) the second  mode. As depicted in Fig. \ref{fig:entropy_2nd_3rd_1_234}(a), we find that unlike the previous case, 
the photon-subtracted state possesses more entanglement in the $1:234$ bipartition than that of the photon-added state.  The ordering remains unchanged if one takes the third mode as player and consider entanglement in the $ 1:234$ split (see Fig. \ref{fig:entropy_2nd_3rd_1_234}(b)). Moreover, we observe that the amount of entanglement decreases in this scenario, compared to the case when the second mode acts as a player.  Note here that if one takes the two-mode squeezed vacuum state as input, it was observed that the bipartite entanglement content of the photon-subtracted state is always lower than that of the photon-added state. 

\newpage

%

\subsubsection{Bipartition with both player and spectator modes}
\label{SubSubSec:2partybipartition}
We still restrict ourselves to the case of a single player. But  
we now  move to the situations in which entanglement of a four-mode state is studied  by considering a bipartition  in which both sides of the split contain two modes, namely $12:34$ and $13:24$.
\begin{figure}[h]
\begin{center}
 \includegraphics[width=0.95\columnwidth,keepaspectratio,angle=0]{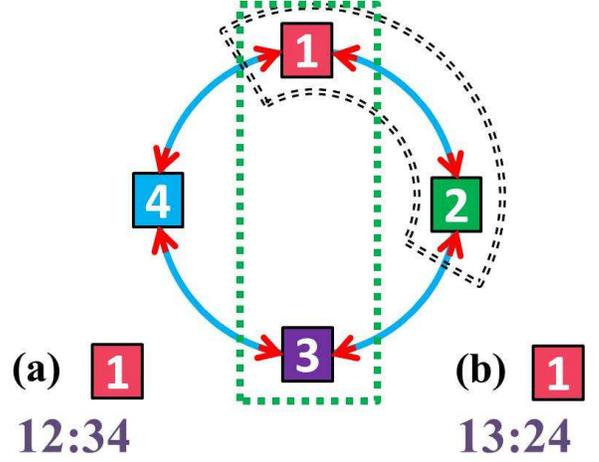}
 \end{center}
\caption{(Color online) Schematic diagram of two different blocks, when a single mode, specifically the first mode, acts as a player.}
\label{fig:Depict_2}
 \end{figure}
The other split between modes, i.e., $14:23$, reflects a  similar behavior, due to the symmetry of the four-mode state. In these two scenarios, photons are added or subtracted  in the first mode, as shown in 
Fig. \ref{fig:Depict_2}, and no photons are added or subtracted, in the other spectator modes.

 To study entanglement of $|\psi^{add}_{m_1}\rangle$ ($|\psi^{sub}_{m_1}\rangle$) in the $12:34$ or $13:24$ bipartition, we require the two party reduced density matrices, $\rho_{12,m_1}^{add}$, $\rho_{13,m_1}^{add}$, $\rho_{12,m_1}^{sub}$, and $\rho_{13,m_1}^{sub}$. 
 We have
%
\begin{widetext}
 \begin{eqnarray}\label{Eq:rho_12_m1_add}
\rho_{12,m_1}^{add} &=& \frac{1}{N_{12}^{add}} \sum_{n,n' = 0}^{\infty}\frac{x^{(n+n')/2}}{2^{n+n'}}\sum_{r_1,r_2=0}^{\min\{n,n'\}}\sqrt{{n \choose r_1}{n \choose r_2}} 
\sqrt{{n' \choose r_1}{n' \choose r_2}}\sqrt{\frac{(n+m_1-r_1)!}{(n-r_1)!}}\sqrt{\frac{(n'+m_1-r_1)!}{(n'-r_1)!}} \nonumber \\
&&\hspace{7cm} \times |n+m_1-r_1\rangle _1 |n - r_2\rangle _2 \langle n'+ m_1-r_1|_1 \langle n' - r_2|_2, 
\end{eqnarray}
and 
\begin{eqnarray}\label{Eq:rho_12_m1_sub}
\rho_{12,m_1}^{sub} &=& \frac{1}{N_{12}^{sub}} \sum_{n,n' = m_1}^{\infty}\frac{x^{(n+n')/2}}{2^{n+n'}}\sum_{r_1=0}^{\min\{n,n'\} - m_1}\sum_{r_2=0}^{\min\{n,n'\}} 
\sqrt{{n \choose r_1}{n \choose r_2}} \sqrt{{n' \choose r_1}{n' \choose r_2}}\sqrt{\frac{(n-r_1)!}{(n-m_1-r_1)!}} 
 \sqrt{\frac{(n'-r_1)!}{(n'-m_1-r_1)!}}\nonumber \\ 
  && \hspace{8cm}  \times|n-m_1-r_1\rangle _1 |n - r_2\rangle _2  \langle n'- m_1-r_1|_1 \langle n' - r_2|_2. 
\end{eqnarray}
\end{widetext}

\begin{figure}[h]
\begin{center}
 \includegraphics[width=1.05\columnwidth,keepaspectratio,angle=0]{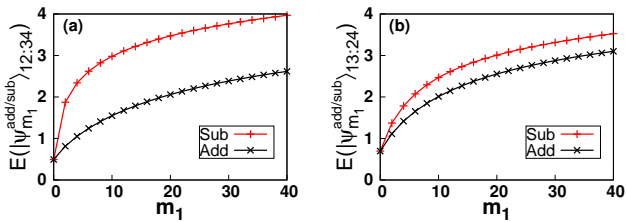}
 \caption{(Color online) Plots of entanglements of photon-added and -subtracted states in the $12:34$  (a) and $13:24$ (b) bipartitions with $m_1$. }
 \label{fig:entropy_1st_12_34__13_24}
 \end{center}
 \end{figure}

Note that in the previous cases, where one partition contains only a single mode, we required single-site density matrices to calculate the entanglement, and  they  are always diagonal in the number basis. The same  is not the case for two-site density matrices. 
Similarly, one can find out the reduced density matrices of $\rho^{add}_{13,m_1}$ and $\rho^{sub}_{13,m_1}$. In both the scenarios, we observe that entanglement increases against the  number of photons added, $m_1$ and same is true for subtraction of photons (see Fig. \ref{fig:entropy_1st_12_34__13_24}). Moreover, as observed in the previous case with the smallest partition consisting of the spectator mode, photon-subtracted state contains higher entanglement in the $12:34$ as well as $13:24$ partitions than that of the corresponding photon-added state. See Figs. \ref{fig:entropy_1st_12_34__13_24}(a) and \ref{fig:entropy_1st_12_34__13_24}(b).


\begin{figure}[h]
\begin{center}
 \includegraphics[width=1\columnwidth,keepaspectratio,angle=0]{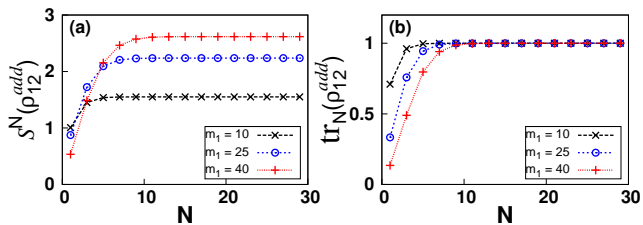}
 \end{center}
\caption{(Color online) Plot of convergence of von Neumann entropy,   ${\cal S}^N(\rho^{add}_{12,m_1})$,  in $(a)$ and $\text{tr}_N(\rho^{add}_{12,m_1})$ in $(b)$ against $N$ which is the maximum value of $n$ and $n'$. We choose three different values of $m_1$, viz. $m_1 = 10, 25, 40$. We find that for example, for $m_1 = 40$, trace goes to unity  and the entropy (entanglement) converges for $N \geq 10$. }
\label{fig:Entropy_tr_with_N}
 \end{figure}

We briefly mention here the method used to calculate $S(\rho_{12, m_1}^{add})$, and the other local entropies. The von Neumann entropy of $\rho_{12,m_1}^{add}$ can be obtained if one can diagonalize the infinite dimensional matrix, given in Eq. (\ref{Eq:rho_12_m1_add}). To calculate it, for fixed $m_1$, we have to truncate the summation upto a large value of $n$ and $n'$, say $N$ for both, and calculate its trace, i.e.,  $\text{tr}_N(\rho^{add}_{12,m_1})$, as well as von Neumann entropy, $S^N(\rho^{add}_{12,m_1})$. We then choose, $2N$ as maximum of $n$ and $n'$ and obtain the quantities. When the difference between $S^N(\rho^{add}_{12,m_1})$ and $ S^{2N}(\rho^{add}_{12,m_1})$ is of the order of $10^{-6}$, we take $S^N(\rho^{add}_{12,m_1})$ as the actual entropy. In Fig \ref{fig:Entropy_tr_with_N}, for a fix value of $m_1$, we plot $S^N(\rho_{12}^{add})$ and $\text{tr}_N(\rho_{12}^{add})$ with the variation of $N$. With the increase of $m_1$, we observe that we require higher values of $N$. However the figure shows both the quantities converge when $N \geq 10$, irrespective of the value  of $m_1$. When we compute entropy or LN, we always carry out a similar scaling analysis for choosing $N$.

\subsection{Behavior of entanglement of photon-added and -subtracted states with two player modes}
In this section, keeping the four-mode squeezed vacuum state as the input state, we increase the number of players from one to two modes, and hence the possibilities of choosing the player modes with nontrivial bipartition grows substantially. 
For a fixed bipartition, we investigate the nature of entanglement by changing the modes in which photons are added or subtracted. 
 Upto now, we have shown that the entanglement content of the resulting state after subtracting photons is either equal or higher than that of the photon-added states. Let us now investigate whether such situation persist when two modes are players.

\begin{figure}[h]
\begin{center}
 \includegraphics[width=0.95\columnwidth,keepaspectratio,angle=0]{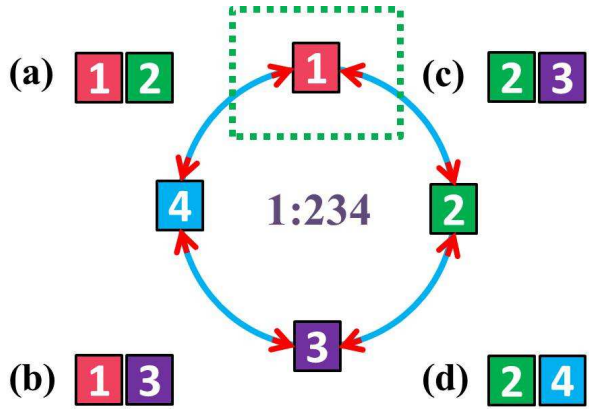}
 \end{center}
\caption{(Color online) Schematic diagram of  four non-trivial possibilities of choosing  two modes as  players in the $1:234$ bipartition. Other choices can be shown as repetitions  due to the symmetry of the FM state.}
\label{fig:Depict_3}
 \end{figure}
 
\subsubsection{One part of the bipartite split contains a single mode}

We begin by concentrating on the entanglement of the FM  state after addition (subtraction) of photons in the  $1:234$ bipartition.
 In this scenario, there are four possibilities for adding and subtracting photons.
  As shown in Fig. \ref{fig:Depict_3}, the modes that act as players are as follows: $(a)$ the first and the second mode, $(b)$ the first and the third mode, $(c)$ the second and the third mode, and $(d)$ the second and the fourth mode. Other possibilities can be reduced to any one of the above four cases due to the symmetry in the four-mode squeezed state. 
Moreover, it can be shown that the entanglement pattern of cases (a) and (b) are qualitatively similar while cases (c) and (d) are analogous and hence the entanglement features will be studied in pairs.

 

\begin{figure}[h]
\begin{center}
 \includegraphics[width=1\columnwidth,keepaspectratio,angle=0]{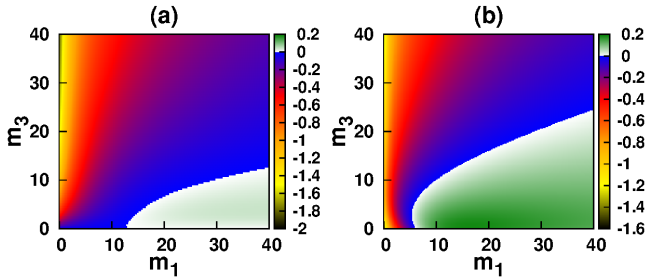}
\caption{(Color online) Behavior of $\delta_1^E (m_1 , m_3 )$ against $m_1$ (horizontal axes) and
$m_3$ (vertical axes). Panels (a) and (b) correspond to inactive and active spectator modes respectively. In Fig. (b) we add (subtract) $m_2 = 5$ photons in (from) the second mode. If both second and fourth modes are active spectators, the region of ${\delta}_1(m_1,m_3)$ increases in $(m_1,m_3)$-plane. For example if we choose $m_2 = $, and $m_4 = $, the region for which $\delta_1^E (m_1 , m_3 ) > 0$ increases. }   
\label{fig:AddminusSub_13_0_m2} 
 \end{center}
 \end{figure}

{\bf Cases} $(a)$ and $(b)$: 
We now consider the situation where either the first and the second modes act as players or the first and the  third modes are players. We calculate the $\delta^E_1(m_1,m_i)$ ($i \neq  1$), when no photons are added and subtracted from the spectator modes.
 We observe that there exists a region for which $\delta_1^E(m_1,m_i) > 0$, which is in contrast with the case when one mode was player in the preceding subsection (see Fig. \ref{fig:AddminusSub_13_0_m2}(a)).
  As seen from the figure, for moderate values of $m_1$, the boundary between  the positive and negative regions  is almost a straight line and hence we can find the slope of the straight line which can help to study these cases quantitatively. We find that for high values of $m_1$, the slope of $\delta_1^E(m_1,m_3) = 0$ is approximately $0.28$, which is small compared to the slope of $\delta_1^E(m_1,m_2) = 0$, which is $0.64$. Moreover, we notice that max $[\delta_1^E(m_1,m_3)] = 0.2 < \max \, \,[\delta_1^E(m_1,m_2)] = 0.4$ while minimum value of $\delta_1^E(m_1,m_3)  ( = -2.0)$ is smaller than that of $\delta_1^E(m_1,m_2) ( = -1.6)$, in the regions surveyed.  Therefore, we can conclude that to create maximal entanglement in this scenario, photon addition is advantageous when one adds photons in the first and the second modes compared to the case of  $m_1$ and $m_3$ being players (with  $m_1 \gg m_i, i=2,3$). 
  
   In both the cases, spectator modes play an important role in the behavior of entanglement in the $1:234$ bipartition. As depicted in Fig. \ref{fig:AddminusSub_13_0_m2} (b), entanglement in the photon-added state can be increased by adding photons in the spectator modes. For example, when $m_{2(4)} = 5$, $\delta_1^E(m_1,m_3)$ against $m_1$ and $m_3$ is depicted in Fig.  \ref{fig:AddminusSub_13_0_m2}(b).
 Quantitative comparison can be made between Figs. \ref{fig:AddminusSub_13_0_m2} (a) and \ref{fig:AddminusSub_13_0_m2} (b). In particular, for $m_1 \gg m_3$, the region  with $\delta_1^E(m_1,m_3) > 0$ when no photons are added (subtracted) in the spectator modes can be calculated. In this limit, we assume that the boundary is a straight line and hence the area is the area of a quadrilateral. Let us call the area as $\Delta_0$. In this case, we calculate the area of the quadrilateral when $m_1 \geq 25$ and $m_1 \leq 40$, and we find $\Delta_0 \approx$ $160$. After adding (subtracting) $5$ photons in the second or fourth modes, we find that the area, $\Delta_5$, of the corresponding quadrilateral increases and  $\Delta_5 \approx$ $253$. 

Behavior of entanglement in the $1:234$ split for cases (c) and (d) are almost identical with the previous cases. The only difference is that entanglement of the subtracted state is always better than that of the added state when spectator modes are inactive. The picture changes, i.e. entanglement of the photon-added states starts increasing faster than the photon-subtracted states, like in the preceding cases, when fixed numbers of photons are added (subtracted) in the spectator mode(s).
 

\subsubsection{Bipartition containing equal number of modes}

\begin{figure}[t]
\begin{center}
 \includegraphics[width=0.95\columnwidth,keepaspectratio,angle=0]{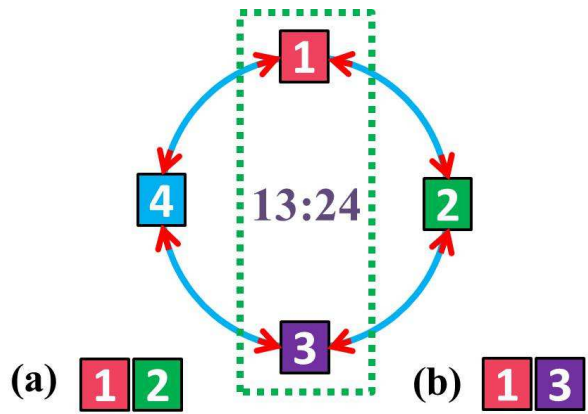}
 \end{center}
\caption{(Color online) Distinct scenarios of two player modes in the $13:24$ split. There are two possibilities -- (a) first and second as players, (b) first and third as players. }
\label{fig:Depict_4}
 \end{figure}

We will now consider the case where, we still keep two modes as players but we now divide four-modes into two blocks consisting of two modes instead of one mode in the preceding discussion. In this case, the  two nontrivial bipartitions are  $12:34$ and $13:24$. Let us first concentrate on the bipartition $13:24$. In this case, the symmetry of the FMSV state after addition or subtraction of arbitrary number of photons in all the modes, given in Eqs.   (\ref{Eq:Photon_Add_def}) and (\ref{Eq:Photon_Sub_def}), ensures that there are only two nontrivial situations in  the case of two player modes (see Fig. \ref{fig:Depict_4}). They are -- $(a)$ when the  players are the first and the second modes, and $(b)$ when first and third modes act as players. Cases (a) and (b) show similar entanglement behavior like previous cases, when one part of the bipartition contains a single mode, and hence we only discuss the situation when two spectator modes are active, which have not been analyzed before. 

{\bf Case} $(a)$: The reduced density matrix of the first and the third mode, for the photon-added state, is given by

\begin{eqnarray}
\rho_{13,\{m_i\}}^{add} &=& \text{tr}_{24}(|\psi_{\{m_i\}}^{add}\rangle \langle \psi_{\{m_i\}}^{add}|)\nonumber \\
&=& \sum_{n=0}^{\infty}\sum_{r_1 = 0}^{n} a_{n,r_1,q} |n+m_1-r_1\rangle _1 |m_3+r_1\rangle _3 \nonumber \\
&& \hspace{6em} \langle n+m_1-q|_1\langle m_3 + q|_3,
\end{eqnarray}  
where we write
\begin{eqnarray}
 a_{n,r_1,q}= \frac{1}{N_{13}^{add}}\frac{x^n}{2^n}\sum_{r_2 = 0}^{n} \Big( {n \choose r_1} {n \choose q}\Big)^{1/2} {n \choose r_2} \nonumber \\
  \Big( \frac{(n+m_1-r_1)!}{(n-r_1)!}\frac{(n+m_1-q)!}{(n-q)!} \frac{(m_3+r_1)!}{r_1!}\frac{(m_3+q)!}{(q)!}\Big)^{1/2}
  \nonumber \\
\frac{(n+m_2-r_2)!}{(n-r_2)!}\frac{(m_4 + r_2)!}{r_2!}. \hspace{2em}
\end{eqnarray}
 Similarly, one can also find the two party reduced density matrix, $\rho_{13,\{m_i\}}^{sub}$, for photon subtraction  by tracing out the second and fourth modes in Eq. (\ref{Eq:Photon_Sub_def}).



If the first and second modes act as players,
 we find that subtraction is always better than addition for arbitrary values of $m_1$ and $m_2$. This case is similar to the case with a single mode being player and cases with second and third modes or second and fourth modes being players.
\begin{figure}[t]
\begin{center}
 \includegraphics[width=1\columnwidth,keepaspectratio,angle=0]{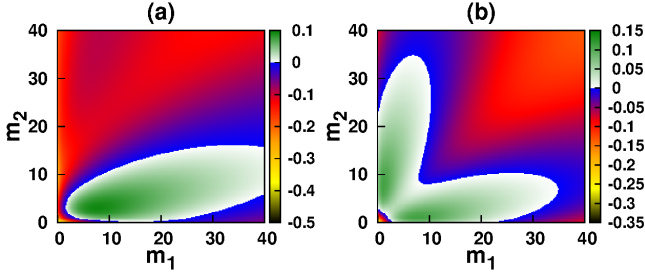} 
 \caption{(Color online) Role of spectator modes in $\delta_{13}^E(m_1, m_2)$. In (a), $m_3 = 10$ and $m_4 = 0$, while in (b), $m_3 = m_4 = 5$. We see that spectator modes help to enhance entanglement in the photon-added state.}
\label{fig:AddminusSub_13_24_12_m3m4} 
 \end{center}
 \end{figure}
To show once more that spectators play a fundamental role in interchanging the entanglement property for photon addition and subtraction, we elaborate the analysis in two scenarios -- (1) when a fixed number of photons are added  (subtracted) in a single spectator mode, a positive region emerges, which indicates that the quantum correlation in the $13:24$ bipartition is greater for photon addition than that for subtraction, as already seen before. An interesting point to note here is that a positive region appears for small values of $m_2$ and almost for all values of $m_1$. This is probably due to the fact that we add photons in the third mode which belongs to the same block as the first mode. (2) When both the spectator modes are active,  the positive region can be seen in both the axis due to symmetry present in the FM state, as depicted in Fig. \ref{fig:AddminusSub_13_24_12_m3m4}(b).

\begin{figure}[t]
\begin{center}
 \includegraphics[width=0.95\columnwidth,keepaspectratio,angle=0]{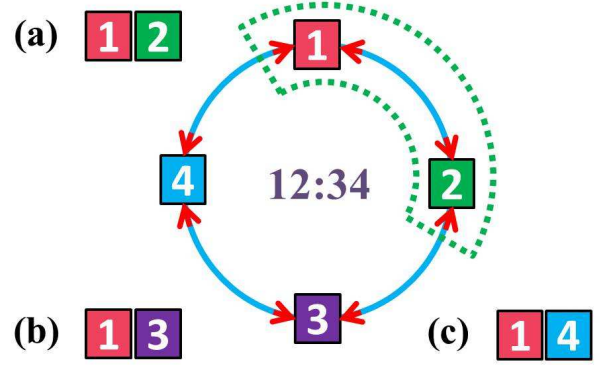}
 \end{center}
\caption{(Color online) Schematic diagram of choices of two player modes in the $12:34$ split.}
\label{fig:Depict_5}
 \end{figure}

Finally, we concentrate on a nontrivial partition, the  $12:34$ cut (see Fig. \ref{fig:Depict_5}). From the perspective of entanglement, this partition is unique. In this scenario, there are three ways to choose the players. We find that  
with and without participation of spectator modes, entanglement of photon subtraction  always higher or equal to that of the photon addition which makes this situation exclusive from others.

\section{Comparison of logarithmic negativity between two-mode and four-mode states}
\label{Sec:LN_Comparison}
Upto now, we have considered an FMSV state as input and have  compared the behavior of entanglement between photon-added and -subtracted states as well as entanglement of an output state in different bipartitions having different player and spectator modes. In this section,  our aim is to make comparison between the output state obtained from the TMSV state after adding or subtracting photons and the two mode state obtained from the FMSV state. To perform such comparison, we discard two modes from the four-mode state and calculate the LN of the two mode reduced state, which we then compare with the LN of the photon-added (subtracted) state that is obtained  from the TMSV state as the input \cite{Cerf_Photon_Add_sub}.
In case of the TM state, the output state, after adding (subtracting) photons, still remains pure and hence LN can be calculated analytically \cite{LN_anlytical}. However, for the FM case, the output state is mixed which is obtained 
by discarding two modes and we adopt the same mechanism as we have done to calculate von Neumann entropy of reduced density matrices, described in Sec. \ref{SubSubSec:2partybipartition}. In particular, we evaluate LN as well as  trace for large $n = N$, and then by increasing $N$, 
we check whether  trace goes to unity upto six decimal points. We truncate the system when trace has already converged to unity, upto six decimal points.  
 
 In the TMSV case, photons can be added to either of the modes or to both the modes. On the other hand, there are several scenarios for the four-mode states. 
 If there is a single player, either one of the mode of the output state can act as player or none of the modes of the output state is the player.
 In case of two players, $(i)$ two players can be the two modes of the output state, $(ii)$ one mode of the output state can be a player, or $(iii)$ the discarded modes can be the player  modes.

Before considering the FMSV state, let us first consider the TMSV state as input.
Note that the nature of LN qualitatively matches with the von Neumann entropy of the reduced density matrix.  
As shown in \cite{Cerf_Photon_Add_sub}, when single mode acts as player, the LN for photon addition coincide with the subtraction,
which is also the case for the von Neumann entropy. If both the modes act as players,
photon addition is always beneficial for entanglement than the photon subtraction \cite{Cerf_Photon_Add_sub}.

\begin{figure}[t]
\begin{center}
 \includegraphics[width=0.95\columnwidth,keepaspectratio,angle=270]{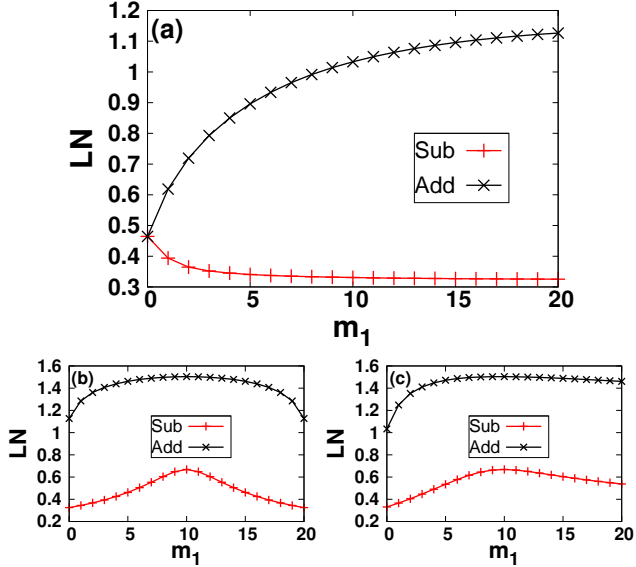} 
 \caption{(Color online) LN betwen  the first and second modes obtained from the FM output state. (a) First mode is player. (b)
 First and second modes are players with $m_1 + m_2 = 20$. (c) First mode is player while the second one is spectator with $m_2 = 10.$}
\label{fig:Log_neg_12_m_1} 
 \end{center}
 \end{figure}

In case of a single player or two players in the FM state, if the output state contains the player mode(s), then the reduced two-mode state obtained from the photon-added state has higher LN than that of the photon-subtracted state. Hence, the behavior of LN of the output state from TM and FM state are identical. As we have shown, this is not the case if we consider the behavior of entanglement of pure four-mode output state in bipartitions. Fig. \ref{fig:Log_neg_12_m_1} depicts the behavior of LN of the two-mode reduced state from the four-mode output state when the first mode acts as  player as well as both the modes of the two mode state are players. In all these situations, no  photons are added (subtracted)  in the spectator modes. We observe that when there is a single player e.g., the first mode of the reduced state, entanglement increases (decreases) monotonically, if photons are added (subtracted). However, such monotonicity with respect to the number of photons added (subtracted) is lost if photons are added (subtracted) in both the modes with total number of photons being fixed as shown in  Fig. \ref{fig:Log_neg_12_m_1}(b).
 A similar qualitative feature in entanglement  is seen when the first mode acts as player while second mode is a spectator having fixed finite number of photons  (see Fig. \ref{fig:Log_neg_12_m_1}(c)). We find that the bipartite entanglement reaches its maximum with respect to $m_1$, when equal number of photons are added (subtracted) in both the modes i.e. $m_1 = m_2$, in Fig  \ref{fig:Log_neg_12_m_1}(b) and $m_1 \approx m_2$ in Fig.  \ref{fig:Log_neg_12_m_1}(c).

\begin{figure}[t]
\begin{center}
 \includegraphics[width=0.39\columnwidth,keepaspectratio,angle=270]{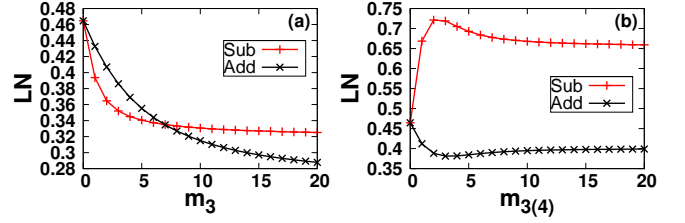} 
 \caption{(Color online) (a) The nature of LN of the first and second mode against the number of photons added (subtracted) in (from) the third mode which has been traced out. (b) LN of the same state in which third as well as fourth modes act as players. The equal number of photons are added (subtracted) in (from) both the modes.}
\label{fig:Log_neg_12_m_3} 
 \end{center}
 \end{figure}
 
Lastly, we consider the scenario, when we add and subtract photons in the discarded modes i.e., in the third and fourth modes,  and we find LN between the first and the second modes, which are spectators. LN of the output state decreases if one of the discarded modes act as a player. For example, by taking third mode as player, we plot LN of the first and the second mode with $m_3$ in Fig. \ref{fig:Log_neg_12_m_3}(a).
  Unlike previous cases, LN of the photon-subtracted state is higher than that of the added state when $m_3 \geq 9$ which can never be observed for the TM case. 
  LN of the photon-subtracted state is more pronounced than that of the added one if both the discarded modes act as players. The same number of photons are added (subtracted) in (from) both the spectator modes, i.e. $m_3 = m_4$, as shown in  Fig. \ref{fig:Log_neg_12_m_3}(b) in which LN($\rho_{34}^{sub}) \geq \text{LN}(\rho_{34}^{add})$.

\section{Non-classicality measure of the photon-added/-subtracted FM state}
\label{Sec:NonG_Comparison}

As mentioned in the introduction, the photon addition and subtraction is one of the ways to create a non-Gaussian state. In this section, we quantify the departure of the photon-added  (-subtracted) FMSV state from Gaussianity, as a function of added (subtracted) photons from the player modes, which was 
introduced in Sec. \ref{Sec:Non_G}.


Since the photon-added (-subtracted) FM state is in a pure state, the second term of $\delta_{NG}(\rho)$, given in Eq. (\ref{Eq:non_G_def}) vanishes. To calculate $\delta_{NG}(\rho)$, we have to find the covariance matrix of $\rho_G$, which is same as $\rho_{\{m_i\}}^{add/sub} = |\psi\rangle\langle \psi|_{\{m_i\}}^{add/sub}$. It is given by
\renewcommand{\arraystretch}{1.5}
\begin{equation}\label{Eq:covariance_matrix}
\sigma_{\varrho} = \left( {\begin{array}{c c c c}
\langle q_1^2\rangle I & \langle q_1q_2\rangle\sigma_z & \langle q_1q_3\rangle I & \langle q_1q_4\rangle \sigma_z \\
\langle q_1q_2\rangle\sigma_z & \langle q_2^2\rangle I & \langle q_2q_3\rangle \sigma_z & \langle q_2q_4\rangle I\\
\langle q_1q_3\rangle I & \langle q_2q_3\rangle \sigma_z & \langle q_3^2\rangle I & \langle q_3q_4 \rangle\sigma_z\\
\langle q_1q_4\rangle \sigma_z & \langle q_2q_4\rangle I & \langle q_3q_4 \rangle\sigma_z & \langle q_4^2 \rangle I\\
\end{array}} \right),  
\end{equation}
where, $q_i = \hat{a}_i + \hat{a}_i^{\dagger}$, and the expectations are taken over the photon-added and -subtracted FM state, given in Eqs. (\ref{Eq:Photon_Add_def}) and (\ref{Eq:Photon_Sub_def}) [for details, see the Appendix].


The Williamson normal form of Eq. (\ref{Eq:covariance_matrix}) can be evaluated by using the prescription given in \cite{Wolf_Williamson_decomposion}. We numerically calculate the Williamson normal form of the matrix in Eq. (\ref{Eq:covariance_matrix}) for both photon addition and subtraction and calculate the non-Gaussianity, which in this case reduces to $ S(\rho_{G,\{m_i\}}^{add/sub})$.
\begin{figure}[h]
\begin{center}
 \includegraphics[width=1.05\columnwidth,keepaspectratio,angle=0]{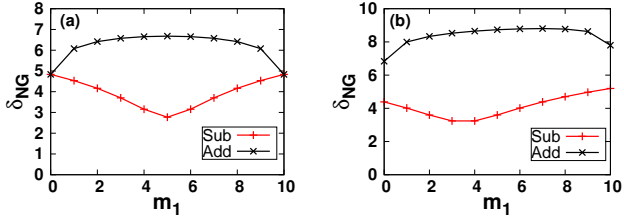}    
 \caption{Behavior of non-Gaussianity measure, $\delta_{NG}$ against $m_1$. (a) First and second modes are players with $m_1 + m_2 = 10$. Spectator modes are ineffective. (b) Spectator modes are active with $m_3 = 4$ and in player modes, $m_1 + m_2 = 10$.}
 \label{fig:Non_G_m1_m2} 
 \end{center}
 \end{figure}
 In all the cases, photon addition leads to a rapid departure of Gaussianity than that of the photon subtraction. We also notice that if among four modes, photons are added only in two modes, then behavior of $\delta_{NG}$ obtained in the FM state and the TM state are  qualitatively similar.
It is clear from the behavior of the non-Gaussianity measure that photon-subtracted state become slowly non-Gaussian as compared to the photon-added state and the behavior remains unchanged irrespective of the choices of the player and the spectator modes (see Fig. \ref{fig:Non_G_m1_m2}).
The rich picture of the role of different modes, captured by entanglement, is not seen by the non-Gaussianity measure and hence indicates that
 there is possibly no direct connection between non-Gaussianity and entanglement content of the output state obtained after photon addition (subtraction) \cite{Cerf_Photon_Add_sub}. 

\section{Conclusion}
\label{Sec:Conclusion}
Photon addition and subtraction constitute useful methods to prepare non-Gaussian states. It has already been established that non-Gaussian states are useful in various quantum mechanical tasks ranging from entanglement distillation to quantum error correction. We have investigated the   entanglement properties of the non-Gaussian states generated by adding or subtracting photons in Gaussian states. In case of two mode states, entanglement of photon-added states are known to be equal or higher than that of the photon-subtracted ones.   

We have shown that this is not the case when one increases the number of modes. We found that for four-mode states, the trend of entanglement distribution in different bipartitions of the photon-added (-subtracted) states is much richer than that in the two-mode states. Specifically, we showed that there exists a scenario, in which multimode entanglement  content of the photon-subtracted state is always higher than that of the corresponding photon-added one. The results remained unchanged even if one discarded two modes from the four-mode output state. Moreover, we showed that the picture that emerges from entanglement of the output state does not match with the behavior in the same states of distance-based non-Gaussianity measures. Upto now, it was known that among addition and subtraction, addition is more beneficial. But our work shows that photon subtraction can also be advantageous if we consider a state of a higher number of modes.

\section{Acknowledgement}
RP acknowledges the INSPIRE-faculty position at Harish-Chandra Research Institute (HRI) from the Department of Science and Technology, Government of India.  

\section*{Appendix}
The expectations used in Eq. (\ref{Eq:covariance_matrix}), for the calculation of non-Gaussianity, taken over the photon-added and -subtracted states are given below.
\begin{equation}
\langle q_1^2\rangle ^{add} = 1 + 2 m_1 + 2\sum_{n,r_1,r_2} (p^{\{m_i\}}_{n,r_1,r_2})^2(n-r_1),
\end{equation}
\begin{equation}
\langle q_2^2\rangle ^{add} = 1 + 2 m_2 + 2\sum_{n,r_1,r_2} (p^{\{m_i\}}_{n,r_1,r_2})^2(n-r_2),
\end{equation}
\begin{equation}
\langle q_3^2\rangle ^{add} = 1 + 2 m_3 + 2\sum_{n,r_1,r_2} (p^{\{m_i\}}_{n,r_1,r_2})^2 r_1,
\end{equation}
\begin{equation}
\langle q_4^2\rangle ^{add} = 1 + 2 m_4 + 2\sum_{n,r_1,r_2} (p^{\{m_i\}}_{n,r_1,r_2})^2 r_2,
\end{equation}
\begin{eqnarray}
\langle q_1q_2\rangle ^{add} &=& 2\sum_{n,r_1,r_2} p^{\{m_i\}}_{n,r_1,r_2}p^{\{m_i\}}_{n+1,r_1,r_2}\nonumber \\ &&   \hspace{-2em} \times \sqrt{(n+m_1-r_1+1)(n+m_2-r_2+1)},
\end{eqnarray}
\begin{eqnarray}
\langle q_1q_3\rangle ^{add} &=& 2\sum_{n,r_2}\sum_{r_1=0}^{n-1} p^{\{m_i\}}_{n,r_1,r_2}p^{\{m_i\}}_{n,r_1+1,r_2}\nonumber \\ &&\times \sqrt{(n+m_1-r_1)(m_3+r_1+1)},
\end{eqnarray}
\begin{eqnarray}
\langle q_1q_4\rangle ^{add} &=& 2\sum_{n,r_1,r_2} p^{\{m_i\}}_{n,r_1,r_2}p^{\{m_i\}}_{n+1,r_1,r_2+1}\nonumber \\ && \times \sqrt{(n+m_1-r_1+1)(m_4+r_2+1)},
\end{eqnarray}
\begin{eqnarray}
\langle q_2q_3\rangle ^{add} &=& 2\sum_{n,r_1,r_2} p^{\{m_i\}}_{n,r_1,r_2}p^{\{m_i\}}_{n+1,r_1+1,r_2}\nonumber \\ && \times \sqrt{(n+m_2-r_2+1)(m_3+r_1+1)},
\end{eqnarray}
\begin{eqnarray}
\langle q_2q_4\rangle ^{add} &=& 2\sum_{n,r_1}\sum_{r_2 = 0}^{n-1} p^{\{m_i\}}_{n,r_1,r_2}p^{\{m_i\}}_{n,r_1,r_2+1}\nonumber \\ && \times \sqrt{(n+m_2-r_2)(m_4+r_2+1)},
\end{eqnarray}
\begin{eqnarray}
\langle q_3q_4\rangle ^{add} &=& 2\sum_{n,r_1,r_2} p^{\{m_i\}}_{n,r_1,r_2}p^{\{m_i\}}_{n+1,r_1+1,r_2+1}\nonumber \\ && \times \sqrt{(m_3+r_1+1)(m_4+r_2+1)},
\end{eqnarray}
where $\sum_{n,r_1,r_2}$ in the photon-added states, is the short form of $\sum_{n = 0}\sum_{r_1 = }^{n}\sum_{r_2 =0}^{n}$. Further, for the photon-subtracted states, 
\begin{equation}
\langle q_1^2\rangle ^{sub} = 1 - 2 m_1 + 2\sum_{n,r_1,r_2} (q^{\{m_i\}}_{n,r_1,r_2})^2(n-r_1),
\end{equation}
\begin{equation}
\langle q_2^2\rangle ^{sub} = 1 - 2 m_2 + 2\sum_{n,r_1,r_2} (q^{\{m_i\}}_{n,r_1,r_2})^2(n-r_2),
\end{equation}
\begin{equation}
\langle q_3^2\rangle ^{sub} = 1 - 2 m_3 + 2\sum_{n,r_1,r_2} (q^{\{m_i\}}_{n,r_1,r_2})^2 r_1,
\end{equation}
\begin{equation}
\langle q_4^2\rangle ^{sub} = 1 - 2 m_4 + 2\sum_{n,r_1,r_2} (q^{\{m_i\}}_{n,r_1,r_2})^2 r_2,
\end{equation}
\begin{eqnarray}
\langle q_1q_2\rangle ^{sub} &=& 2\sum_{n,r_1,r_2} q^{\{m_i\}}_{n,r_1,r_2}q^{\{m_i\}}_{n+1,r_1,r_2}\nonumber \\ &&   \hspace{-2em} \times \sqrt{(n-m_1-r_1+1)(n-m_2-r_2+1)},
\end{eqnarray}
\begin{eqnarray}
\langle q_1q_3\rangle ^{sub} &=& 2\sum_{n,r_2}\sum_{r_1=m_3}^{n-m_1-1} q^{\{m_i\}}_{n,r_1,r_2}p^{\{m_i\}}_{n,r_1+1,r_2}\nonumber \\ &&\times \sqrt{(n-m_1-r_1)(r_1-m_3+1)},
\end{eqnarray}
\begin{eqnarray}
\langle q_1q_4\rangle ^{sub} &=& 2\sum_{n,r_1,r_2} q^{\{m_i\}}_{n,r_1,r_2}q^{\{m_i\}}_{n+1,r_1,r_2+1}\nonumber \\ && \times \sqrt{(n-m_1-r_1+1)(r_2 -m_4 +1)},
\end{eqnarray}
\begin{eqnarray}
\langle q_2q_3\rangle ^{sub} &=& 2\sum_{n,r_1,r_2} q^{\{m_i\}}_{n,r_1,r_2}q^{\{m_i\}}_{n+1,r_1+1,r_2}\nonumber \\ && \times \sqrt{(n-m_2-r_2+1)(r_1-m_3+1)},
\end{eqnarray}
\begin{eqnarray}
\langle q_2q_4\rangle ^{sub} &=& 2\sum_{n,r_1}\sum_{r_2 = m_4}^{n-m_2-1} q^{\{m_i\}}_{n,r_1,r_2}q^{\{m_i\}}_{n,r_1,r_2+1}\nonumber \\ && \times \sqrt{(n-m_2-r_2)(r_2 - m_4+1)},
\end{eqnarray}
\begin{eqnarray}
\langle q_3q_4\rangle ^{sub} &=& 2\sum_{n,r_1,r_2} q^{\{m_i\}}_{n,r_1,r_2}q^{\{m_i\}}_{n+1,r_1+1,r_2+1}\nonumber \\ && \times \sqrt{(r_1-m_3+1)(r_2-m_4+1)},
\end{eqnarray}
where $\sum_{n,r_1,r_2}$, is the short form of $\sum_{n = M}\sum_{r_1 = m_3}^{n-m_1}\sum_{r_2 = m_4}^{n-m_2}$, and $M = \max\{m_1+m_3,m_2+m_4\}$.


\begin{thebibliography}{999}

\bibitem{HHHH_RMP} R. Horodecki, P. Horodecki, M. Horodecki, and K. Horodecki, Rev. Mod. Phys. {\bf 81}, 865 (2009).

\bibitem{QCryp} M. {\.Z}ukowski, A. Zeilinger, M. A. Horne, and H. Weinfurter, Acta Phys. Pol. A {\bf 93}, 187 (1998); 
M. Hillery, V. Bu\^zek, and A. Berthiaume, Phys. Rev. A {\bf 59}, 1829 (1999); 
A. Karlsson, M. Koashi, and N. Imoto, {\em ibid.} {\bf 59}, 162 (1999);  
R. Cleve, D. Gottesman, and H.-K. Lo, Phys. Rev. Lett. {\bf 83}, 648 (1999); 
D. Gottesman, Phys. Rev. A {\bf 61}, 042311 (2000); 
W. Tittel, H. Zbinden, and N. Gisin, {\em ibid.} {\bf 63}, 042301 (2001); 
N. Gisin, G. Ribordy, W. Tittel, and H. Zbinden, Rev. Mod. Phys. {\bf 74}, 145 (2002); 
A. Sen(De), U. Sen, and  M. {\.Z}ukowski, Phys. Rev. A {\bf 68}, 032309 (2003); 
S. K. Singh and R. Srikanth, Phys. Rev. A {\bf 71}, 012328 (2005); 
Z. -J. Zhang, Y. Li, and Z. -X. Man, {\em ibid.} {\bf 71}, 044301 (2005); 
K. Chen and H. -K. Lo, Quantum Inf. Comput. {\bf 7}, 689 (2007); 
D. Markham and B. C. Sanders, Phys. Rev. A {\bf 78}, 042309 (2008); 
R. Demkowicz-Dobrza\'nski, A. Sen(De), U. Sen, and M. Lewenstein, {\em ibid.} {\bf 80}, 012311 (2009); 
A. Marin and D. Markham, {\em ibid.} {\bf 88}, 042332 (2013).

\bibitem{DistriDC} D. Bru{\ss}, G. M. D'Ariano, M. Lewenstein, C. Macchiavello, A. Sen(De), and U. Sen, Phys. Rev. Lett. {\bf 93}, 210501 (2004).

\bibitem{Q_state_distri} 
J. I. Cirac, P. Zoller, H. J. Kimble, and H. Mabuchi, Phys. Rev. Lett. {\bf 78}, 3221 (1997); 
S. Bose, Contemp. Phys. {\bf 48}, 13 (2007); 
G. M. Nikolopoulos and I. Jex, {\em Quantum State Transfer and Network Engineering} (Springer, Heidelberg, 2013).

\bibitem{cluster_comp} R. Raussendorf and H. J. Briegel, Phys. Rev. Lett. {\bf 86}, 5188 (2001); F. Meier, J. Levy, and D. Loss, {\em ibid.} {\bf 90}, 047901 (2003); R. Raussendorf, D. E. Browne, and H. J. Briegel, Phys. Rev. A {\bf 68}, 022312 (2003); M. A. Nielsen, Phys. Rev. Lett. {\bf 93}, 040503 (2004); P. Walther, K. J. Resch, T. Rudolph, E. Schenck, H. Weinfurter, V. Vedral, M. Aspelmeyer, and A. Zeilinger, Nature {\bf 434}, 169 (2005); M. A. Nielsen, Rep. Math. Phys. {\bf 57}, 147 (2006);
H. J. Briegel, D. E. Browne, W. D\" ur, R. Raussendorf, and M. V. den Nest, Nat. Phys.
{\bf 5}, 19 (2009).


\bibitem{Pan10photon}W. -B. Gao, C. -Y. Lu, X. -C. Yao, P. Xu, O. G\"uhne, A. Goebel, Y. -A. Chen, C. -Z. Peng, Z. -B. Chen, and J. -W. Pan, Nat. Phys. \textbf{6}, 331 (2010);
 T. E. Northup and R. Blatt, Nat. Photonics {\bf 8}, 356 (2014).


\bibitem{ION} D. Leibfried, R. Blatt, C. Monroe, and D. Wineland, Rev. Mod. Phys. {\bf 75}, 281 (2003); H. H\" affner, C. Roos, and R. Blatt, Phys. Rep. {\bf 469}, 155 (2008); T. Monz, P. Schindler, J. T. Barreiro, M. Chwalla, D. Nigg, W. A. Coish, M. Harlander, W. H\"ansel, M. Hennrich, and R. Blatt, Phys. Rev. Lett. {\bf 106}, 130506 (2011); J. T. Barreiro, J. -D. Bancal, P. Schindler, D. Nigg, M. Hennrich, T. Monz, N. Gisin and R. Blatt, Nat. Phys. \textbf{9}, 559 (2013).


\bibitem{NMR} L. M. K. Vandersypen and I. L. Chuang, Rev. Mod. Phys. {\bf 76}, 1037 (2005); C. Negrevergne, T. S. Mahesh, C. A. Ryan, M. Ditty, F. Cyr-Racine, W. Power, N. Boulant, T. Havel, D. G. Cory, and R. Laflamme, Phys. Rev. Lett. \textbf{96}, 170501 (2006).

\bibitem{NVcenter}
W. L. Yang, Z. Q. Yin, Z. Y. Xu, M. Feng, and J. F. Du, Appl. Phys. Lett. {\bf 96}, 241113 (2010);
P. C. Maurer, G. Kucsko, C. Latta, L. Jiang, N. Y. Yao, S. D. Bennett, F. Pastawsk, D. Hunger, N. Chisholm, M. Markham, D. J. Twitchen, J. I. Cirac, and M. D. Lukin, Science {\bf 336}, 1283 (2012); 
K. Nemoto, M. Trupke, S. J. Devitt, A. M. Stephens, B. Scharfenberger, K. Buczak,
T. N\" obauer, M. S. Everitt, J. Schmiedmayer, and W. J. Munro, Phys. Rev. X {\bf 4}, 031022 (2014).



\bibitem{EPR_paper} A. Einstein, B. Podolsky, and N. Rosen, Phys. Rev. {\bf 47}, 777 (1935).


%
\bibitem{TeleBennett} C. H. Bennett, G. Brassard, C. Cr\'epeau, R. Jozsa, A. Peres, and W. K. Wootters, Phys. Rev. Lett. {\bf 70}, 1895 (1993).
%
%

\bibitem{DC1992} C. H. Bennett, and S. J. Wiesner, Phys. Rev. Lett. {\bf 69}, 2881 (1992).


\bibitem{Braun_tele} S. L. Braunstein, L. H. Kimble, Phys. Rev. Lett. {\bf 80}, 869 (1998).


\bibitem{Holevo_Capacity_CV} A. S. Holevo, M. Sohma, O. Hirota, Phys. Rev. A {\bf 59}, 1820 (1999).

\bibitem{Multi_ent_mea_CV} S. L. Braunstein, P. V. Loock, Rev. Mod. Phys. {\bf 77}, 513 (2005).
\bibitem{CV_reali} P. Kok, W. J. Munro, K. Nemoto, T. C. Ralph, J. P. Dowling, and G. J. Milburn,
Rev. Mod. Phys. {\bf 79}, 135 (2007); J. Eisert, and M. B. Plenio,  	Int. J. Quant. Inf. {\bf 1}, 479 (2003); F. Dell’Anno, S. De Siena, F. Illuminati, Phys. Rep. {\bf 428}, 53 (2006).

\bibitem{Ent_dist} J. Eisert, S. Scheel, and M. B. Plenio, Phys. Rev. Lett. {\bf 89}, 137903 (2002); G. Giedke and J. I. Cirac,  Phys. Rev. A {\bf 66}, 032316 (2002). 

\bibitem{Measurement_qun_com} M. Ohliger, K. Kieling, and J. Eisert, Phys. Rev. A
{\bf 82}, 042336 (2010); M. Ohliger and J. Eisert, Phys. Rev. A {\bf 85}, 062318 (2012).

\bibitem{NonG_tele} F. Dell’Anno, S. De Siena, L. Albano, and F. Illuminati, Phys. Rev. A {\bf 76}, 022301 (2007); F. Dell’Anno, S. De Siena, and F. Illuminati, Phys. Rev. A {\bf 81}, 012333 (2010).

\bibitem{Non_G_error_cor} S. Lloyd and S. L. Braunstein, Phys. Rev. Lett. {\bf 82}, 1784 (1999).


\bibitem{lab_state} G. S. Agarwal and K. Tara, Phys. Rev. A {\bf 43}, 492 (1991); A. Zavatta, A. Viciani, and M. Bellini, Science {\bf 306}, 660 (2004);  J. Wenger, R. Tualle-Brouri, and P. Grangier, Phys. Rev. Lett. {\bf 92}, 153601 (2004); K. Wakui, H. Takahashi, A. Furusawa, and M. Sasaki, Opt. Exp. {\bf 15}, 3568 (2007);  A. I. Lvovsky and S. A. Babichev, Phys. Rev. A {\bf 66}, 011801 (2002); V. D’Auria, C. de Lisio, A. Porzio, S. Solimeno, J. Anwar, and M. G. A. Paris, Phys. Rev. A {\bf 81}, 033846 (2010); V. D’Auria, A. Chiummo, M. De Laurentis, A. Porzio, and S. Solimeno, Opt. Exp. {\bf 13}, 948 (2005).

\bibitem{Wigner} E. Wigner, Phys. Rev. {\bf 40}, 749 (1932).

\bibitem{R_prabhu_non_c} A. R. Usha Devi, R. Prabhu, and M. S. Uma, Eur. Phys. J. D {\bf 40}, 133 (2006).

\bibitem{Cerf_Photon_Add_sub} T. Opatrn\' y, G. Kurizki, D. -G. Welsch Phys.Rev. A {\bf 61},  032302 (2000); S. Olivares, M. G. A. Paris, R. Bonifacio  Phys. Rev. A {\bf  67}, 032314 (2003); S. Y. Lee, S. W. Ji, H. J. Kim, and H. Nha
Phys. Rev. A {\bf 84}, 012302 (2011); C. Navarrete-Benlloch, R. Garc\'ia-Patr\'on, J. H. Shapiro, and N. J. Cerf, Phys. Rev. A, {\bf 86}, 012328 (2012).

\bibitem{lab_demo}  M. Dakna, L. Kn\" oll, and D. -G. Welsch, Eur. Phys. J. D {\bf 3}, 295 (1998);
V. Parigi, A. Zavatta, M. S. Kim, and M. Bellini, Science {\bf 317}, 1890 (2007); A. Ourjoumtsev, A. Dantan, R. Tualle-Brouri, and P. Grangier Phys. Rev. Lett. {\bf 98}, 030502 (2007); M. Kim, J. Phys. B {\bf 41}, 133001 (2008);
M. Bellini and A. Zavatta, Prog. Optics {\bf 55}, 41 (2010);
 Y. Kurochkin, A. S. Prasad, and A. I. Lvovsky, Phys. Rev. Lett. {\bf 112}, 070402 (2014).
 
\bibitem{Ma_Rhodes} X. Ma and W. Rhodes, Phys. Rev. A {\bf 41}, 4625 (1990); L. Hu and H. Fan, Europhys. Lett. {\bf 85}, 60001 (2009).    
 
\bibitem{Optics_book} M. O. Scully and M. S. Zubairy, \emph{Quantum Optics} (Cambridge University Press, 1997).

\bibitem{Non_G_Genoni}M. G. Genoni, M. G. A. Paris, K. Banaszek, Phys. Rev. A. {\bf 78}, 060303(R) (2008).
\bibitem{Quantify_non_G_Marco_Genoni} M. G. Genoni, M. G. A. Paris, Phys. Rev. A. {\bf 82}, 052341 (2010).
\bibitem{Quantify_non_G_Marian}  P. Marian and T. A. Marian, Phys. Rev. A {\bf 88}, 012322 (2013).


\bibitem{Ent_entropy}  C. H. Bennett, H. J. Bernstein, S. Popescu, and B. Schumacher, Phys. Rev. A {\bf 53}, 2046 (1996).

\bibitem{distillable}  C. H. Bennett, D. P. DiVincenzo, J. A. Smolin, and W. K. Wootters, Phys. Rev. A {\bf 54}, 3824 (1996);
E. M. Rains, {\em ibid.}  {\bf 60}, 173 (1999); {\em ibid.} {\bf 60}, 179 (1999); 
P. M. Hayden, M. Horodecki, and B. M. Terhal, J. Phys. A: Math. Gen. {\bf 34}, 6891 (2001).

\bibitem{concEof} S. Hill and W. K. Wootters, Phys. Rev. Lett. {\bf 78}, 5022 (1997); 
	W.K. Wootters, {\em ibid.} {\bf 80}, 2245 (1998).
	
\bibitem{LN} J. Lee, M. S. Kim, Y. J. Park, and S. Lee, J. Mod. Opt. {\bf 47}, 2151 (2000); 
G. Vidal and R.F. Werner, Phys. Rev. A {\bf 65}, 032314 (2002);  
M. B. Plenio, Phys. Rev. Lett. {\bf 95}, 090503 (2005)

\bibitem{Plenio_RMP} J. Eisert, M. Cramer, and M. B. Plenio, Rev. Mod. Phys. {\bf 82}, 277 (2010).


\bibitem{PTranspose} A. Peres, Phys. Rev. Lett. {\bf 77}, 1413 (1996); M. Horodecki, P. Horodecki, and R. Horodecki, Phys. Lett. A {\bf 223}, 1 (1996).


\bibitem{LN_anlytical} A. Kitagawa, M. Takeoka, M. Sasaki, A. Chefles, Phys. Rev. A. {\bf 73}, 042310 (2006). 

\bibitem{Wolf_Williamson_decomposion} M. M. Wolf, Phys. Rev. Lett. {\bf 100}, 070505 (2008).



\end{thebibliography}
 \end{document}